\documentclass[prb,aps,twocolumn,superscriptaddress,showpacs]{revtex4}

\usepackage{amsmath,amssymb,graphicx}
\usepackage{pxfonts}
\usepackage{graphicx}

\begin{document}

\title{Optically Controlled Excitonic Transistor}

\author{P.~Andreakou}
\affiliation{Department of Physics, University of California at San Diego, La Jolla, CA 92093-0319, USA}
\affiliation{Laboratoire Charles Coulomb, Universit{\' e} Montpellier 2, CNRS, UMR 5221, F-34095 Montpellier, France}

\author{S.V.~Poltavtsev}
\affiliation{Department of Physics, University of California at San Diego, La Jolla, CA 92093-0319, USA}
\affiliation{Spin Optics Laboratory, St.~Petersburg State University, St.~Petersburg, Russia}

\author{J.R.~Leonard}
\affiliation{Department of Physics, University of California at San Diego, La Jolla, CA 92093-0319, USA}

\author{E.V.~Calman}
\affiliation{Department of Physics, University of California at San Diego, La Jolla, CA 92093-0319, USA}

\author{M.~Remeika}
\affiliation{Department of Physics, University of California at San Diego, La Jolla, CA 92093-0319, USA}

\author{Y.Y.~Kuznetsova} \email{yuliyakuzn@gmail.com}
\affiliation{Department of Physics, University of California at San Diego, La Jolla, CA 92093-0319, USA}

\author{L.V.~Butov}
\affiliation{Department of Physics, University of California at San Diego, La Jolla, CA 92093-0319, USA}

\author{J.~Wilkes}
\affiliation{Department of Physics and Astronomy, Cardiff University, Cardiff CF24 3AA, United Kingdom}

\author{M.~Hanson} \author{A.C.~Gossard}
\affiliation{Materials Department, University of California at Santa Barbara, Santa Barbara, CA 93106-5050, USA}

\begin{abstract}
\noindent Optical control of exciton fluxes is realized for indirect excitons in a crossed-ramp excitonic device. The device demonstrates experimental proof of principle for all-optical excitonic transistors with a high ratio between the excitonic signal at the optical drain and the excitonic signal due to the optical gate. The device also demonstrates experimental proof of principle for all-optical excitonic routers.
\end{abstract}

\maketitle

Excitonic devices control potential energy landscapes for excitons, exciton fluxes, emission rates, and other characteristics of excitons. Excitonic devices are used both for studies of basic properties of excitons and for development of excitonic signal processing. The potential advantages of excitonic signal processing include compact footprint and high interconnection speed \cite{Baldo09}. The development of excitonic devices, where exciton fluxes are controlled, is mainly concentrated on indirect excitons in coupled quantum wells (CQW) \cite{Hagn95, Gartner06, High07, High08, Grosso09, Remeika09, High09, Vogele09, Kuznetsova10, Cohen11, Winbow11, Schinner11, Remeika12, Leonard12} and exciton-polaritons in microcavities \cite{Amo10, Gao12, Ballarini12, Nguyen13, Sturn13}.

An indirect exciton is a bound pair of an electron and a hole in spatially separated QW layers (Fig.~1a). The overlap of electron and hole wavefunctions for indirect excitons can be engineered by the structure design and applied voltage so that lifetimes of indirect excitons can exceed those of regular excitons by orders of magnitude. Long lifetimes allow indirect excitons to travel over sufficiently large distances to accommodate devices. Indirect exciton energy can be controlled by voltage: an electric field $F_z$ perpendicular to the QW plane results in the exciton energy shift $e d F_z$, where $ed$ is the built-in dipole moment of indirect excitons ($d$ is close to the distance between the QW centers for a CQW structure) \cite{Miller85}. Control of indirect excitons transport by voltage was demonstrated in electrostatically created in-plane potential landscapes $E(x,y) = - e d F_z(x,y)$ in excitonic ramps \cite{Hagn95, Gartner06, Leonard12}, circuit devices \cite{High07, High08, Grosso09, Kuznetsova10}, narrow channels \cite{Grosso09, Vogele09, Cohen11}, traps \cite{High09, Schinner11}, lattices \cite{Remeika09, Remeika12}, and conveyers \cite{Winbow11}.

\begin{figure*}[htbp]
\centering
\includegraphics[width=17cm]{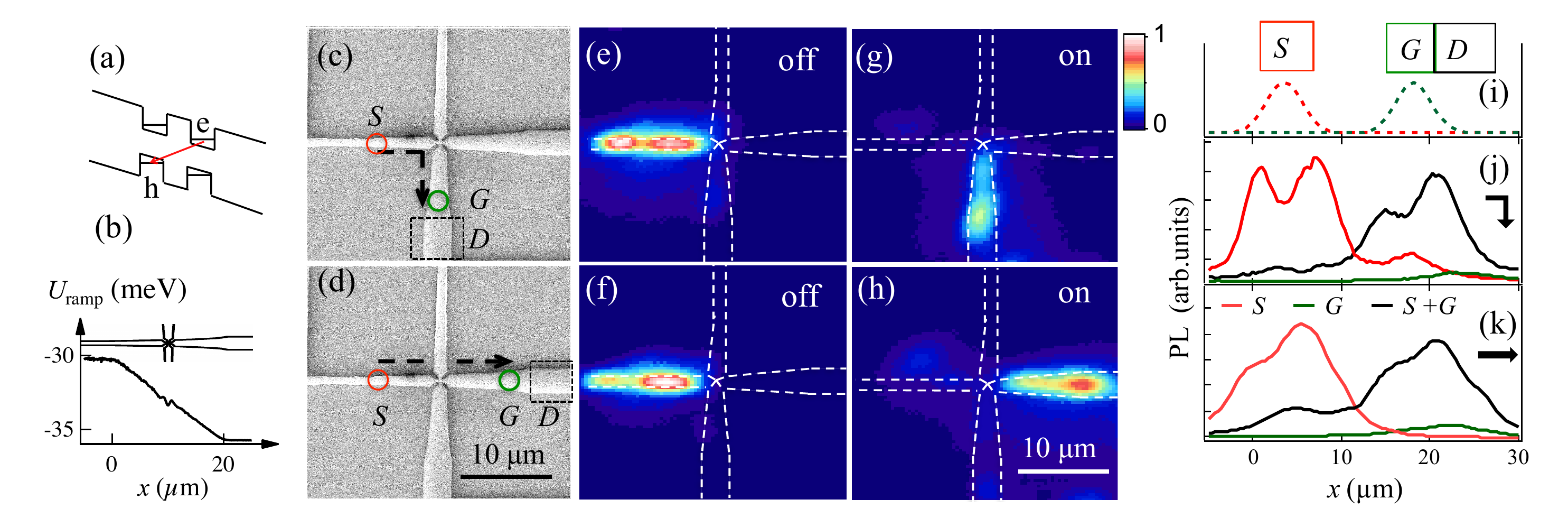}
\caption{(a) CQW band diagram. e, electron, h, hole. (b) Simulated potential for indirect excitons in the crossed-ramp device. (c,d) SEM images of the electrode. The shaped electrode forms the crossing ramps for indirect excitons. Red and green circles indicate the excitation spots of the source ($S$) and gate ($G$) beam, respectively. Arrows indicate the turned-path (c) or straight-path (d) operation of the excitonic transistor. (e-h) Images of the exciton emission in (e,f) off and (g,h) on states for the turned-path (e,g) and straight-path (f,h) transistor operation. The power of the source beam $P_S=0.5~\mu$W. The power of the gate beam $P_G=0$ (e,f) and $0.2~\mu$W (g,h). (j,k) Emission intensity of indirect excitons along the exciton flux for the turned-path (j) and straight-path (k) transistor operation in off state (red, $P_S= 0.5~\mu$W, $P_G=0$), in on state (black, $P_S=0.5~\mu$W, $P_G=0.2~\mu$W), and when only the gate beam is on (green, $P_S=0, P_G=0.2~\mu$W). (i) Spatial profiles of the source (red) and gate (green) excitation beam spots.}
\end{figure*}

In this work, we study indirect excitons in a crossed-ramp excitonic device. A potential energy gradient - a ramp - is created by a shaped electrode at constant voltage with no energy-dissipating voltage gradient \cite{Leonard12}. The design utilizes the ability to control exciton energy by electrode density \cite{Kuznetsova10a}. An electrode on the sample surface - the top electrode (Fig.~1c,d) - is shaped so that a voltage applied between it and a homogeneous bottom electrode creates two crossing ramps. In each of these ramps, narrowing of the top electrode reduces $F_z$ due to field divergence near the electrode edges and, as a result, increases the exciton energy. The electrode shape (Fig.~1c,d) is designed to obtain a constant potential energy gradient for indirect excitons in the CQW along each of the crossing ramps (Fig.~1b). At the crossing point, the electrodes are narrowed to compensate for the otherwise increased electrode density and thus to keep the exciton energy linear along both ramps \cite{Kuznetsova10a}. Each of the two ramps is surrounded by flat-energy channels where the electrode width and, in turn, the energy of indirect excitons, is constant (Fig.~1b-d). The parameters of the structure and experimental details are presented in supplementary materials.

The crossed-ramp excitonic device demonstrates experimental proof of principle for all-optical excitonic transistors with optical input, output, and control gate using indirect excitons as the operation medium. Photons transform into excitons at the optical input (source) and travel to the optical output (drain) due to the ramp potential. The output signal of the exciton emission in the drain region is controlled by a gate beam.

Figure~1e,f shows the images of the emission of indirect excitons along the turned-path and straight-path of the crossed-ramp device when only the source beam is on. The corresponding spatial profiles of the indirect exciton emission intensity $I(x)$ are presented in Fig.~1j,k. The profile of the source beam is shown in Fig.~1i. The emission patterns show enhanced emission intensity around the excitation spot due to the inner-ring effect studied earlier \cite{Ivanov06, Hammack09}. The inner ring was explained in terms of exciton transport and cooling: the heating of the exciton gas by laser excitation reduces the occupation of low-energy optically active exciton states, in turn reducing the exciton emission intensity in the excitation spot. When excitons travel away from the excitation spot, they thermalize to the lattice temperature, and the occupation of low-energy optically active exciton states increases, in turn increasing the exciton emission intensity and forming photoluminescence ring around the excitation spot \cite{Ivanov06, Hammack09}. Emission patterns also show the enhanced emission intensity of indirect excitons at the lower energy side of the ramps due to exciton transport along the potential energy gradient.

The regime when only the source beam is on refers to off state of the excitonic transistor. The output signal of the transistor given by the emission of indirect excitons in the drain region is weak (Fig.~1j,k). The exciton signal in the drain region is controlled by an optical gate. The positions of the gate beam and drain region are shown in Fig.~1c and Fig.~1d for the turned-path or straight-path operation of the excitonic transistor, respectively. Figure~1j,k shows that turning on the gate beam strongly increases the exciton signal at the drain, switching the excitonic transistor to the on state. Figure~2a,b shows that the on/off ratio of the output intensity in the drain region reaches two orders of magnitude. Furthermore, even a weak gate beam generating a weak exciton signal (green lines in Fig.~1j,k) can strongly increase the output in the drain. Figure~2c,d shows that the ratio between the excitonic signal at the optical drain and the excitonic signal due to the optical gate $I_{D-{\rm on}}/I_G$ reaches an order of magnitude. We note that the excitonic signal due to the optical gate is generally smaller than the gate signal since not all of the gate photons transform to the emitted photons due to losses, therefore, in general $I_{D-{\rm on}}/I_G$ is not equal to the transistor gain.

The crossed-ramp excitonic device also demonstrates an experimental proof of principle of all-optical excitonic routers. In the absence of the gate beam, the output signal of exciton emission is low in both ramps after the crossing point (Fig.~1e,f). Positioning the gate beam on one or another ramp of the crossed-ramp device determines the path where the output signal is directed (Fig.~1c-h). Even a weak gate beam generating a weak exciton signal can route a much stronger output signal of exciton emission (Fig.~1j,k).

\begin{figure}[htbp]
\centering
\includegraphics[width=7cm]{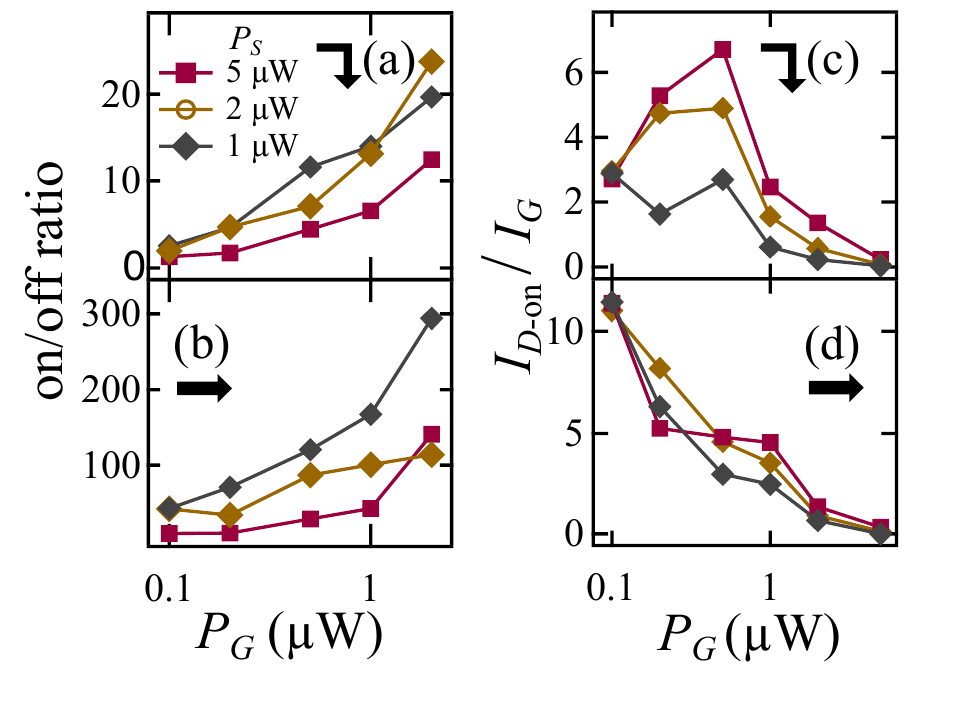}
\caption{(a,b) The contrast ratio of the output intensity integrated over the drain region. (c,d) The ratio of the exciton emission intensity integrated over the drain region in on state to the exciton emission intensity integrated over the entire device when only the gate beam is on. The data are shown for the turned-path (a,c) and straight-path (b,d) operation of the excitonic transistor.}
\end{figure}

Below we present the theoretical model for the exciton transport in a crossed-ramp device. Indirect exciton transport in a 1D potential energy channel was modeled using the following non-linear transport equation:
\begin{equation}
\nabla \left[ D \nabla n + \mu n \nabla (u_0 n + U_{\rm ramp}) \right] - n/\tau + \Lambda_S + \Lambda_G = 0.
\end{equation}
This equation was solved for the steady-state distribution of indirect excitons, $n$ with $\nabla=\partial/\partial x$ (the coordinate $x$ follows the path of the exciton flux along the energy channel and ramp). Diffusion and drift fluxes are given by the first and second terms in square brackets, respectively. The exciton diffusion coefficient $D$ and mobility $\mu$ are related by the generalized Einstein relationship \cite{Ivanov06, Hammack09}, $\mu = D(e^{T_0/T} - 1)/(k_B T_0)$ where $T_0 = (\pi \hbar^2 n)/(2 M k_B)$ is the quantum degeneracy temperature and $M$ is the exciton mass. The drift flux is due to the gradient in the applied ramp potential $U_{\rm ramp}$ (shown in Fig.~3d) and the exciton-exciton interaction. The exciton-exciton interaction was modeled as a repulsive dipolar interaction potential, approximated by $u_0 n$ with $u_0 = 4\pi de^2/\varepsilon_b$, where $\varepsilon_b$ is the GaAs dielectric constant. The effect of the QW disorder potential is included in the diffusion coefficient, $D = D_0\exp[-U_0/(u_0 n + k_B T)]$ \cite{Ivanov06, Hammack09}. Here, $U_0/2 = 0.5\,{\rm meV}$ is the amplitude of the disorder potential and $D_0$ is the diffusion coefficient in the absence of disorder. This description includes (i) increased exciton localization to potential minima for decreasing temperature and (ii) screening of the disorder potential for increasing exciton density \cite{Ivanov06, Hammack09}.

\begin{figure}[htbp]
\centering
\includegraphics[width=5.1cm]{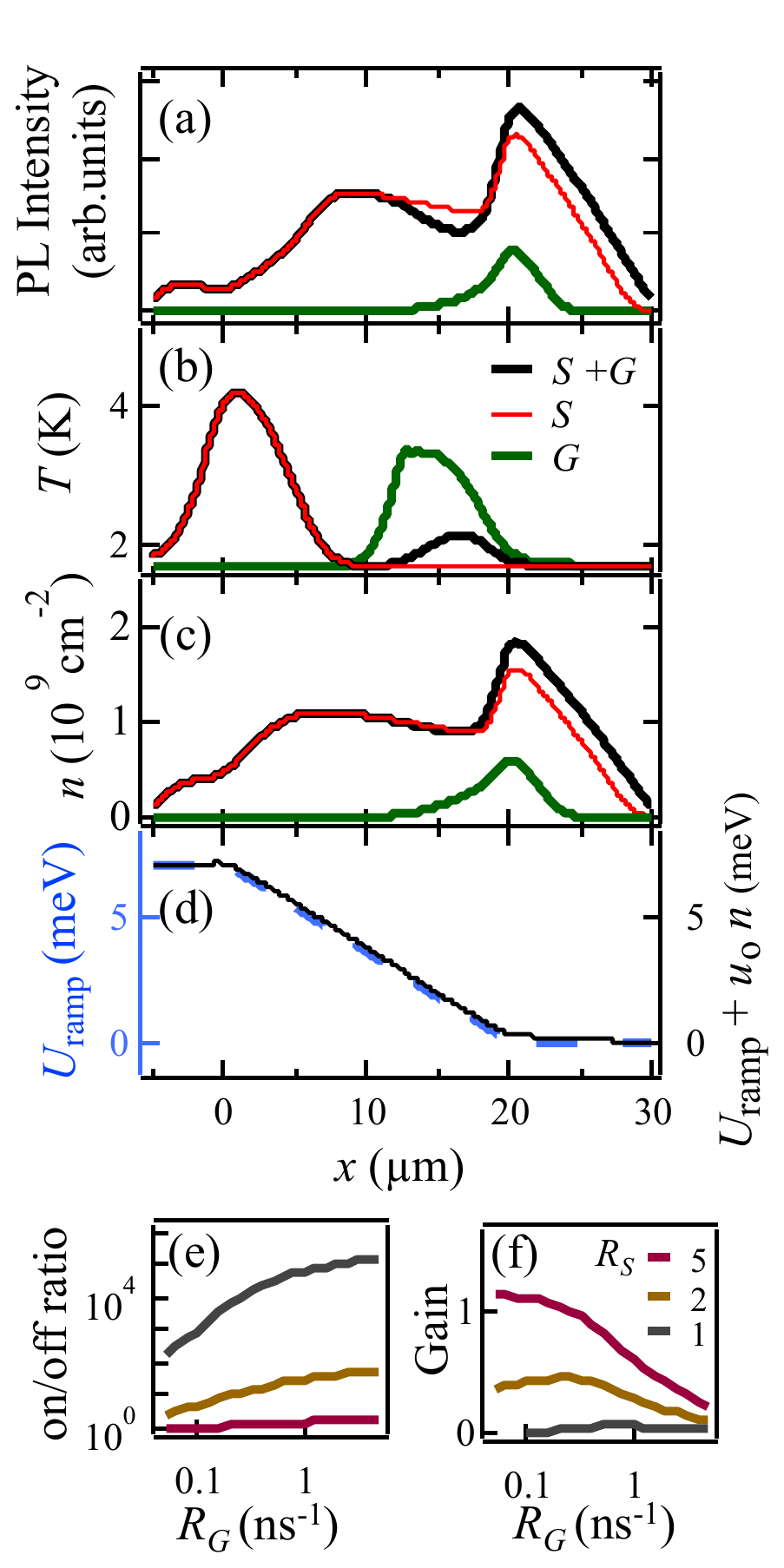}
\caption{Theoretical simulations: Exciton emission intensity (a), temperature (b), and density (c) for $R_S=5$~ns$^{-1}$ and $R_G=0$ (red, off state), $R_S=0$ and $R_G=0.5$~ns$^{-1}$ (green), and $R_S=5$~ns$^{-1}$ and $R_G=0.5$~ns$^{-1}$ (black, on state). (d) Bare ramp potential $U_{ramp}$ (dashed blue) and screened ramp potential $U_{ramp} + u_0 n$ for $R_S=5$~ns$^{-1}$ and $R_G=0.5$~ns$^{-1}$ (black). (e) The contrast ratio of the output intensity in the drain region vs. gate generation rate $R_G$ for different source generation rates $R_S$. (f) The gain of the excitonic transistor. The gain is defined as the ratio of the exciton emission intensity integrated over the drain region in on state to the exciton emission intensity integrated over the entire device when only the gate beam is on.}
\end{figure}

Optical generation of excitons at the source (gate) is given by $\Lambda_{S(G)}$. These have Gaussian profiles with position and FWHM chosen to match the experiment. The total injection rate of excitons is $R_{S(G)} = \delta y \int \Lambda_{S(G)}(x) dx$ where $\delta y \approx 1\,{\rm \mu m}$ is the width of the potential energy channel. The exciton optical lifetime $\tau(T_0,T)$ determines the decay of optically active excitons \cite{Ivanov06, Hammack09}. The spatially varying exciton temperature used in Eq. (1) is found by solving a thermalization equation:
\begin{equation}
S_{\rm phonon}(T_0,T) = S_{\rm pump}(T_0,T,\Lambda_S + \Lambda_G,E_{\rm inc}).
\end{equation}
Here, $S_{\rm phonon}$ accounts for energy relaxation of excitons due to bulk longitudinal acoustic phonon emission. $S_{\rm pump}$ is the heating rate due to the non-resonant laser excitation and is determined by the generation rates, $\Lambda_{S(G)}$ and the excess energy of photoexcited excitons $E_{\rm inc}$. Expressions for $S_{\rm phonon}$, $S_{\rm pump}$, $\tau$, and all other parameters of the model are given in Ref. \cite{Hammack09}.

The results of the simulations are presented in Fig.~3. Figure~3d shows the bare ramp potential $U_{ramp}$ and the ramp potential screened by the exciton-exciton repulsion $U_{ramp} + u_0 n$. Figures 3a, 3b, and 3c show the emission intensity, temperature, and density of indirect excitons. Figures 3e and 3f show the contrast ratio of the output intensity in the drain region and the gain of the excitonic transistor for different source and gate powers.

Within the model, excitons generated in the gate area screen the disorder in the structure due to the repulsive dipolar interactions of indirect excitons. This increases the source exciton transport distance along the ramp, allowing excitons generated by the source laser to reach the drain region. The transport distance is further enhanced by the gate-beam-induced heating of the exciton gas due to the thermal activation of excitons in the disorder potential. The heating also increases the exciton lifetime that increases the density and, in turn, the screening of the disorder potential. The increasing exciton transport distance along the ramp generates the signal in the drain region. The features observed in the model are qualitatively similar to the experimental data, compare Figs.~1j,k with Fig.~3a, Fig.~2a,b with Fig.~3e, and Fig.~2c,d with Fig.~3f. However, the experimentally observed ratio between the excitonic signal at the optical drain to the excitonic signal due to the optical gate is much higher than the gain of the excitonic transistor calculated within the model. This indicates that a model beyond the considered drift/diffusion model should be developed for a quantitative description of the experimental data.

In summary, we report on experimental proof of principle for all-optical excitonic transistors with a high ratio of the excitonic signal at the optical drain to the excitonic signal due to the optical gate and experimental proof of principle for all-optical excitonic routers.

We thank Marc Baldo, Alexey Kavokin, and Masha Vladimirova for discussions. This work was supported by NSF Grant No. 0907349. P.A. was supported by EU ITN INDEX. S.V.P. was supported by a strategic MIT-Skoltech development program and the Russian Ministry of Education and Science. Y.Y.K. was supported by an Intel fellowship. J.W. was supported by EPSRC. Work in Cardiff University was performed using the computational facilities of the ARCCA Division.

\end{document}

\renewcommand{\thefigure}{S\arabic{figure}}

\title{Supplementary Materials for Optically Controlled Excitonic Transistor}

\author{P.~Andreakou}
\affiliation{Department of Physics, University of California at San Diego, La Jolla, CA 92093-0319, USA}
\affiliation{Laboratoire Charles Coulomb, Universit{\' e} Montpellier 2, CNRS, UMR 5221, F-34095 Montpellier, France}

\author{S.V.~Poltavtsev}
\affiliation{Department of Physics, University of California at San Diego, La Jolla, CA 92093-0319, USA}
\affiliation{Spin Optics Laboratory, St.~Petersburg State University, St.~Petersburg, Russia}

\author{J.R.~Leonard}
\affiliation{Department of Physics, University of California at San Diego, La Jolla, CA 92093-0319, USA}

\author{E.V.~Calman}
\affiliation{Department of Physics, University of California at San Diego, La Jolla, CA 92093-0319, USA}

\author{M.~Remeika}
\affiliation{Department of Physics, University of California at San Diego, La Jolla, CA 92093-0319, USA}

\author{Y.Y.~Kuznetsova} \email{yuliyakuzn@gmail.com}
\affiliation{Department of Physics, University of California at San Diego, La Jolla, CA 92093-0319, USA}

\author{L.V.~Butov}
\affiliation{Department of Physics, University of California at San Diego, La Jolla, CA 92093-0319, USA}

\author{J.~Wilkes}
\affiliation{Department of Physics and Astronomy, Cardiff University, Cardiff CF24 3AA, United Kingdom}

\author{M.~Hanson} \author{A.C.~Gossard}
\affiliation{Materials Department, University of California at Santa Barbara, Santa Barbara, CA 93106-5050, USA}

\maketitle

The CQW structure was grown by molecular beam epitaxy. An $n^+$-GaAs layer with $n_{Si}$ = $10^{18}$ cm$^3$ serves as a bottom electrode. A semitransparent top electrode is fabricated by depositing a 100 nm indium tin oxide layer. Two 8 nm GaAs QWs separated by a 4 nm Al$_{0.33}$Ga$_{0.67}$As barrier are positioned 100 nm above the $n^+$-GaAs layer within an undoped $1\,\mu$m thick Al$_{0.33}$Ga$_{0.67}$As layer. Positioning the CQW closer to the homogeneous electrode suppresses the in-plane electric field \cite{Hammack06}, which otherwise can lead to exciton dissociation \cite{Zimmerman97}. Excitons are photoexcited by a 633 nm HeNe laser (data in the main text and Fig.~S2 in supplementary materials) or 787.5 nm Ti:Sapphire laser (Fig.~S1 in supplementary materials). The former wavelength corresponds to the photon energy above the Al$_{0.33}$Ga$_{0.67}$As barrier and the latter corresponds to the photon energy at the direct exciton resonance. Experiments are performed at $T_{bath}=1.6$ K. For the data in the main text, the voltage applied to the top electrode $V_e= - 3$~V.

\begin{figure*}[htbp]
\centering
\includegraphics[width=15cm]{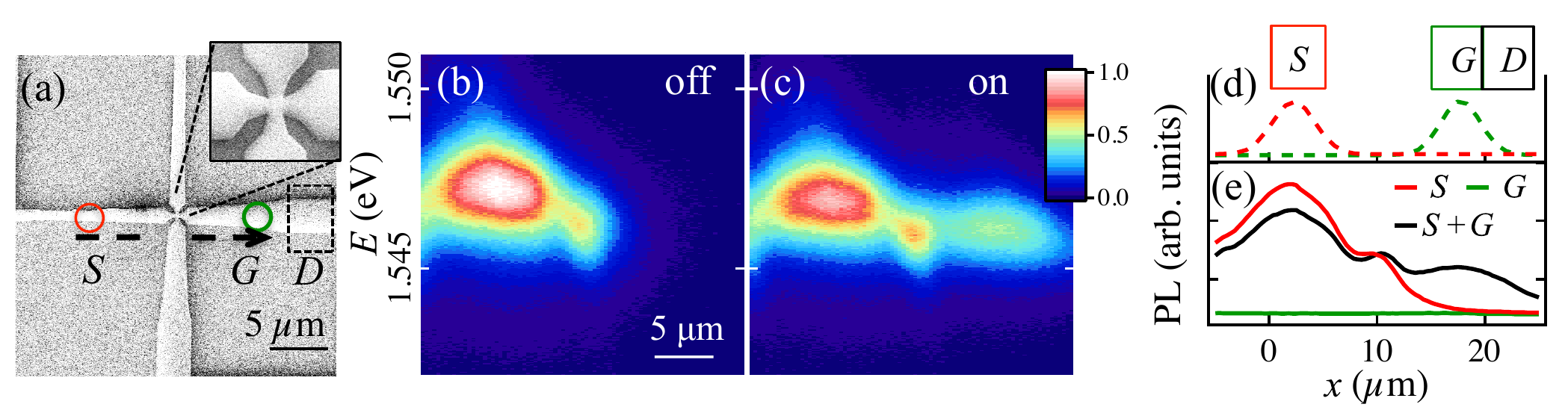}
\caption{(a) SEM images of the crossed-ramp device. Inset: SEM image of the crossing point of the crossed-ramp device. Red and green circles indicate the excitation spots of the source ($S$) and gate ($G$) beam, respectively. The arrow indicates the operation path of the excitonic transistor. (b,c) Energy-resolved images of the exciton emission in (b) off and (c) on states for the crossed-ramp transistor. The power of the source beam $P_S=2~\mu$W (b,c). The power of the gate beam $P_G=0$ (b) and $2~\mu$W (c). (d) Spatial profiles of the source (red) and gate (green) excitation beam spots. (e) Emission intensity of indirect excitons along the exciton flux for the crossed-ramp transistor in off state [red, $P_S= 2~\mu$W, $P_G=0$], in on state [black, $P_S=2~\mu$W, $P_G=1~\mu$W], and when only the gate beam is on [green, $P_S=0$, $P_G=1~\mu$W]. $V_e= - 4$~V.}
\end{figure*}

\begin{figure*}[htbp]
\centering
\includegraphics[width=16cm]{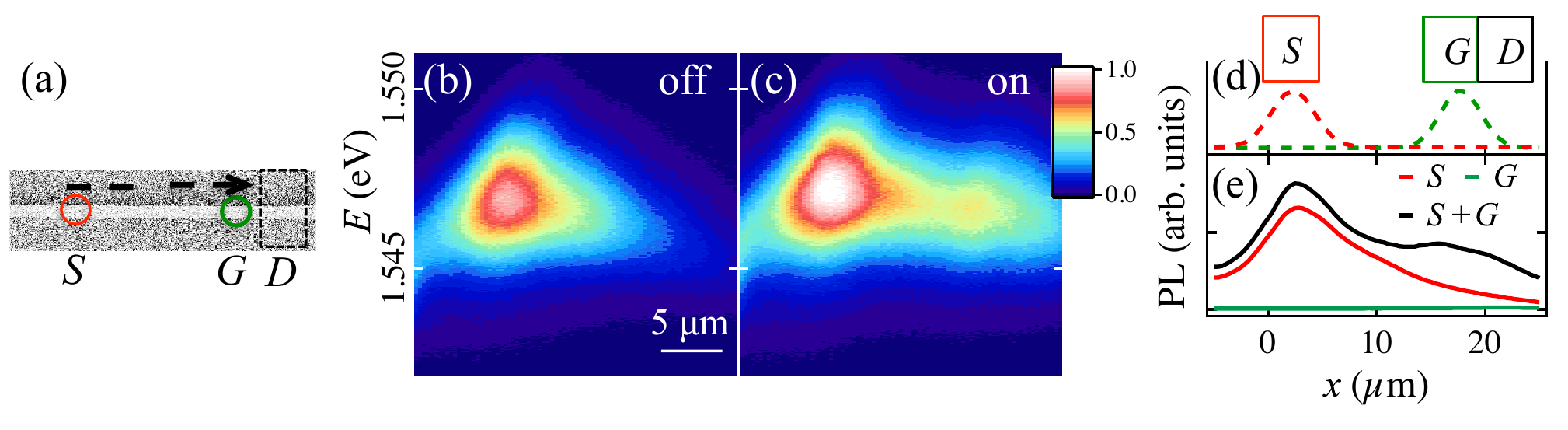}
\caption{(a) SEM images of the planar-electrode device. Red and green circles indicate the excitation spots of the source ($S$) and gate ($G$) beam, respectively. (b,c) Energy-resolved images of the exciton emission in (b) off and (c) on states for the planar-electrode transistor. The power of the source beam $P_S=10~\mu$W (b,c). The power of the gate beam $P_G=0$ (b) and $2~\mu$W (c). (d) Spatial profiles of the source (red) and gate (green) excitation beam spots. (e) Emission intensity of indirect excitons for the planar-electrode transistor in off state [red, $P_S= 10~\mu$W, $P_G=0$], in on state [black, $P_S=10~\mu$W, $P_G=2~\mu$W], and when only the gate beam is on [green, $P_S=0$, $P_G=2~\mu$W]. $V_e= - 4$~V.}
\end{figure*}

Qualitatively similar data are measured for the source and gate beam wavelengths 633 nm (main text) and 787.5 nm (Fig.~S1). Furthermore, qualitatively similar data are measured for the cross-ramp device (main text) and planar-electrode device, which forms a flat-energy channel (Fig.~S2). In all these cases (i) the gate beam strongly increases the exciton signal at the drain region and (ii) even a gate beam, which generates a weak exciton signal, can strongly increase the output at the drain (Figs. 1 and 2 in the main text and Figs. S1 and S2). This indicates that all-optical excitonic transistors can operate at various source and gate beam wavelengths and various electrode geometries.